\documentclass{IEEEtran}

\usepackage{cite}
\ifCLASSINFOpdf
   \usepackage[pdftex]{graphicx}
\else
   \usepackage[dvips]{graphicx}
\fi
\usepackage[cmex10]{amsmath}
\usepackage{amsfonts}
\usepackage{amssymb}

\ifCLASSOPTIONcompsoc
  \usepackage[caption=false,font=normalsize,labelfont=sf,textfont=sf]{subfig}
\else
  \usepackage[caption=false,font=footnotesize]{subfig}
\fi

\usepackage{multicol}
\usepackage{xcolor}
\usepackage{comment}

\newcommand{\bfg}[1]{\boldsymbol{#1}}

\newcommand{\I}{\bar{\bfg \imath}}
\newcommand{\Y}{\bar{\bfg Y}}
\newcommand{\Z}{\bar{\bfg Z}}
\newcommand{\V}{\bar{\bfg v}}

\newcommand{\idir}{\bar{\imath}_{h\rightarrow j}}
\newcommand{\sdir}{\bar{s}_{h\rightarrow j}}
\newcommand{\sdirdev}{\bar{s}_{h\rightarrow j\,d}}
\newcommand{\etadev}[1]{\overline{\eta^{s}}_{h\rightarrow j\,#1}}
\newcommand{\etaidev}[1]{\overline{\eta^{\imath}}_{#1}}

\begin{document}

\title{Coherency among Power System Devices}

\author{
\IEEEauthorblockN{Ignacio Ponce, Rodrigo Bernal and Federico Milano\\}
\IEEEauthorblockA{School of Electrical \& Electronic Engineering \\
University College Dublin,
Dublin, Ireland\\
ignacio.poncearancibia@ucdconnect.ie,
rodrigo.bernal@ucdconnect.ie,
federico.milano@ucd.ie}
\thanks{This work is supported by the Sustainable Energy Authority of Ireland (SEAI) by funding I. Ponce, R. Bernal and F. Milano under project FRESLIPS, Grant No. RDD/00681.
}
\vspace{-0.8cm}}

\maketitle

\begin{abstract}
  The paper proposes a novel general definition of coherency among power system devices of any type.  The proposed approach is thus not limited to synchronous machines.  With this aim, the paper shows that coherency can be formally based on the difference in the complex frequency of the current injections of any two devices electrically connected to the same grid.  The proposed definition is model-agnostic, making it general and suitable for modern power systems composed of a heterogeneous mix of technologies.  The paper also provides a systematic analytical procedure to study the properties that specific device models must satisfy to be coherent.  Time-domain simulations are conducted in three case studies whose results illustrate the ability of our definition to evaluate coherency among any type of device.
\end{abstract}


\section{Introduction}

\subsection{Motivation}

\textit{Coherency} in power systems has been traditionally employed to refer to a similarity in the dynamic response of synchronous machines (SMs) following a disturbance.  For decades, coherency has been successfully applied to conventional power systems for a wide range of applications, primarily in the context of transient stability studies \cite{chow}.
The foundations of coherency analysis are challenged in the case of modern power systems, characterized by the presence of highly heterogeneous devices that introduce more complex dynamics, giving rise to a wide variety of technical challenges \cite{milano2018foundations}.  Furthermore, recent revisions to power system stability definitions \cite{taskforcerev} have made existing coherency theory and applications inadequate.  This paper aims to address this gap by revisiting and generalizing the definition of the concept of coherency among power system devices.

\subsection{Literature review}

Early works provided formal definitions of theoretical coherency as a mathematical condition between state variables of the model of SMs, namely, two machines are coherent if the angular difference between them is constant, also implying their angular speed is equal \cite{dicaprio}.  This definition has been crucial for applying the concept of coherency in power systems, including the aggregation of coherent groups of SMs into a dynamic equivalent unit, reducing the computational effort involved \cite{machowski, podmore2007identification, wu2003coherency, dynred}, control solutions \cite{padhy2013coherency, dorfler2014sparsity, vahidnia2015wide}, and equipment placement \cite{chakrabortty, klein, rios, kamwa2002pmu}, among other applications \cite{koochi2019taxonomy}.

Given the proliferation of converter-interfaced resources, recent works have focused on extending the use of coherency to capture the behavior of these devices and enable their aggregation.  Several examples are on the coherency-based aggregation of wind turbines (WTs), where early works evaluate coherency based on specific internal variables, such as the WT's speed \cite{zhao2009coherency, jin2009dynamic, ni2016model}.  Later works have improved the coherency identification methods to suit different WT type models \cite{iravani1, iravani2}.  The authors in \cite{li2015coherency} presented a coherency-based method to equivalence multilevel converters using virtual synchronous generator control, leveraging the fact they mimic SMs.  Finally, notable works use the Koopman operator for identifying coherency in nonlinear systems, allowing aggregating full-rated converters based on Koopman mode analysis \cite{susuki, budivsic2012applied, chamorro}.  

In the field of signal processing, the coherency between time-domain signals is often formalized based on the cross-correlation function \cite{oppenheim}, while in the frequency domain it is expressed through spectral density functions \cite{random}.  Both definitions require the knowledge of the signals over a given finite time period before evaluating coherence between them.  An example is found in \cite{schimmel}, where the authors define instantaneous phase coherency and use it for signal extraction from ambient noise.  However, this approach requires using the Hilbert transform to construct an equivalent analytical signal, and only uses the instantaneous phase.

Existing literature on coherency analysis is mainly focused on applying the concept for equivalencing a group of devices of a certain type or mix.  These approaches adapt the definition of coherency depending on variables specific to the models under study.  Besides, although early noticed, most works do not take into account the dynamics of the voltage despite their crucial role for complete coherency \cite{1295018}.  Consequently, a general definition of coherency that is model-agnostic, easily measurable, and trackable in the time-domain is still lacking for power system devices.  In this paper we aim at addressing this gap by presenting a novel definition of coherency precisely with those features.

\subsection{Contributions}

The contributions of the paper are twofold:
\begin{itemize}
    \item A general definition of coherency among power system devices, suitable for evaluating the degree of similarity in the dynamic response of any pair of devices 
    --- not only synchronous machines --- connected to the grid.
    \item A systematic procedure for studying the properties that device dynamic models must satisfy to be coherent.
\end{itemize}

\subsection{Paper organization}

The remainder of the paper is organized as follows.  Section \ref{sec:definition} proposes a novel coherency definition and a specific formulation for transient stability studies.  Section \ref{sec:coherency_devices} applies it to common devices to identify specific requirements among models to achieve coherency.  Section \ref{sec:casestudy} demonstrates practical assessment through a series of relevant case studies, and Section \ref{sec:conclusion} concludes the paper and outlines future research opportunities.

\section{A novel definition of coherency}
\label{sec:definition}

Our goal is to provide a novel definition of coherency among power system devices that reflects the degree to which components of a system present a similar dynamic behavior.  Accordingly, two perfectly coherent devices need not be distinct entities, since neither contains additional information beyond what is already known by the other.  They are rather a scaled or shifted manifestation of a unique underlying dynamic.

In order to achieve this goal, we require our definition to meet the following guidelines:
\begin{itemize}
    \item Suitable for evaluating the dynamic behavior of any pair --- or any two clusters --- of devices composing the power system.
    \item Model-agnostic, making it a generalized property regardless of the special features of each device.  
    \item Instantaneous, making it a well-defined property regardless of the time period under study.
\end{itemize}

Considering three-phase balanced power systems, electrical quantities can be conveniently represented using Clarke vectors, which lie in a two-dimensional $(\alpha, \beta)$ space with well-defined magnitude and phase, enabling an instantaneous definition without the need for a Hilbert transform.  Unlike the definition found in \cite{schimmel}, we consider the dynamics of the two components of the Clarke vectors.  As demonstrated later in Section~\ref{sec:casestudy}, phase coherency among devices is insufficient to achieve the points of the list above.  

The full dynamics of a Clarke vector can be conveniently represented using the so-called complex-frequency (CF), originally introduced in \cite{cf}.  The CF is a particular case of the geometric frequency for three-phase systems under balanced conditions \cite{geometricalf}.  Without lack of generality and for the sake of simplicity, our definition is expressed in terms of the CF, but we remark that it can be broadened using the geometric frequency to handle higher-dimensional objects, such as multi-phase systems or unbalanced conditions.

\subsection{General definition}

Consider two Clarke vectors in polar coordinates $\bar{x}(t)=x(t)\,e^{\jmath\,\phi_x(t)}$ and $\bar{y}(t)=y(t)\,e^{\jmath\,\phi_y(t)}$ representing two devices composing the power system.  The dependency on time is hereinafter omitted for simplicity.  The CF of $\bar{x}$ and $\bar{y}$ is denoted as $\bar{\eta}_{\bar{x}}$ and $\bar{\eta}_{\bar{y}}$ respectively, defined as follows:
\begin{equation}
    \bar{\eta}_{\bar{x}}=\frac{\dot{x}}{x}+\jmath\,\dot{\phi}_x\,,\quad  \bar{\eta}_{\bar{y}}=\frac{\dot{y}}{y}+\jmath\,\dot{\phi}_y\,,
\end{equation}
where the dot over a quantity denotes its time derivative.

The instantaneous coherency function $\bar{\epsilon}$ between $\bar{x}$ and $\bar{y}$ is defined as follows:
\begin{equation}\label{eq:gendef}
\boxed{
\bar{\epsilon}(\bar{x}, \bar{y})=\bar{\eta}_{\bar{x}}-\bar{\eta}_{\bar{y}}
}
\end{equation}

Note in case $\bar{x}$ and $\bar{y}$ are perfectly coherent, $\bar{\eta}_{\bar{x}}=\bar{\eta}_{\bar{y}}\Rightarrow\bar{\epsilon}(\bar{x}, \bar{y})=0$.  

What remains is to find which physical quantities $\bar{x}$ and $\bar{y}$ are most meaningful to represent the devices.  The answer to this question depends on the system model.  In the section below, we discuss the typical network model used for transient stability studies.

\subsection{Coherency between shunt-connected devices to an algebraic model of the transmission network}
In this section, we apply the instantaneous coherency definition given by (\ref{eq:gendef}) to a pair of shunt-connected devices in an algebraic transmission network, i.e., the fast dynamics of the network are neglected.  For this model, the current network equations in matrix form are:
\begin{equation}\label{eq:KCL}
    \I=\Y\cdot\V\,,
\end{equation}
where $\Y$ is the admittance matrix of the network, and $\I$ and $\V$ are column vectors containing the net current injections and voltages at every bus of the grid.

In case $\Y$ has full-rank\footnote{Although this condition holds under practical assumptions \cite{mario}, for theoretical cases for which $\rm{det}(\Y) = 0$, our definition of coherency can still be evaluated using (\ref{eq:gendef}) and a proper choice of Clarke vectors.}, (\ref{eq:KCL}) is equivalent to:
\begin{equation}\label{eq:Ohm}
    \V=\Z\cdot\I\,,
\end{equation}
where $\Z=\Y^{-1}$ is the impedance matrix of the network.

Taking the $h$-th equation of (\ref{eq:Ohm}):
\begin{equation}\label{eq:ohm_h}
    \bar{v}_h=\sum_{k\in \mathbb{B}}\bar{z}_{hk}\,\bar{\imath}_k\,,
\end{equation}
where $\mathbb{B}$ is the set of buses of the system.  

Let $\idir$ be the current flowing from bus $h$ to an arbitrary direction $j$, either towards a neighboring bus or drawn by a shunt device at $h$.  We multiply the conjugate of $\idir$ on both sides of (\ref{eq:ohm_h}) to get the complex power flowing from bus $h$ to the direction $j$:
\begin{equation}
    \idir^{*}\,\bar{v}_h=\sdir=\sum_{k\in \mathbb{B}}\idir^{*}\,\bar{z}_{hk}\,\bar{\imath}_k\,.\label{eq:sdir0}
\end{equation}

The net current injected at $k$, i.e., $\bar{\imath}_k$, can be written as the sum of the currents injected by every device connected at $k$:
\begin{equation}\label{eq:ikdev}
    \bar{\imath}_k=\sum_{d\in\mathbb{D}_k}\bar{\imath}_d\,,
\end{equation}
where $\mathbb{D}_k$ is the set of devices connected at bus $k$.  

Using (\ref{eq:ikdev}) in (\ref{eq:sdir0}):
\begin{align}
    \sdir&=\sum_{k\in \mathbb{B}}\sum_{d\in\mathbb{D}_k}\idir^{*}\,\bar{z}_{hk}\,\bar{\imath}_d\,.\label{eq:sdir1}
\end{align}
We denote $b_d$ the bus on which device $d$ is connected.  For instance, $\forall d\in\mathbb{D}_k$, $b_d=k$.  With this notation, (\ref{eq:sdir1}) can be written more compactly as:
\begin{equation}\label{eq:sdir2}
    \sdir=\sum_{d\in\mathbb{D}}\idir^{*}\,\bar{z}_{hb_{d}}\,\bar{\imath}_d\,.
\end{equation}

We define:
\begin{equation}
    \sdirdev:=\idir^{*}\,\bar{z}_{hb_d}\,\bar{\imath}_d\,,
\end{equation}
which represents the contribution of device $d$ to the complex power in bus $h$ towards $j$.

Equation (\ref{eq:sdir2}) can therefore be written as the sum of the individual contributions of all devices in the system:
\begin{equation}\label{eq:superpos}
    \sdir=\sum_{d\in\mathbb{D}}\sdirdev\,.
\end{equation}

From (\ref{eq:superpos}), $\sdirdev$ appears as the representative quantity to evaluate coherency of device $d$ against any other.  At this point, coherency appears to be a property between devices depending on the location and direction of interest ($h\rightarrow j$).  However, we show below that this is not the case.

Having identified the representative Clarke vectors for evaluating coherency, the derivation continues by applying the general definition (\ref{eq:gendef}).  Considering two arbitrary devices $d_{\rm{1}}$ and $d_{\rm{2}}$, the coherency function between them is:
\begin{equation}
\bar{\epsilon}(\bar{s}_{h\rightarrow j\,d_{\rm 1}}, \bar{s}_{h\rightarrow j\,d_{\rm 2}})=\etadev{d_{\rm{1}}}-\etadev{d_{\rm{2}}}\,,
\end{equation}
where:
\begin{equation}
    \etadev{d}=\overline{\eta^{\imath}}_{h\rightarrow j}^{*}+\overline{\eta^{\imath}}_d\,.
\end{equation}

Therefore:
\begin{align}
    \bar{\epsilon}(\bar{s}_{h\rightarrow j\,d_{\rm 1}}, \bar{s}_{h\rightarrow j\,d_{\rm 2}})&=
    \overline{\eta^{\imath}}_{h\rightarrow j}^{*}+\overline{\eta^{\imath}}_{d_{\rm 1}}-\overline{\eta^{\imath}}_{h\rightarrow j}^{*}-\overline{\eta^{\imath}}_{d_{\rm 2}}\,.
\end{align}

Hence:
\begin{equation}
\boxed{
    \bar{\epsilon}(\bar{s}_{h\rightarrow j\,d_{\rm 1}}, \bar{s}_{h\rightarrow j\,d_{\rm 2}})=\overline{\eta^{\imath}}_{d_{\rm 1}}-\overline{\eta^{\imath}}_{d_{\rm 2}}\,.}\label{eq:coherency_function}
\end{equation}

The result above implies that the coherency function between two devices is the same regardless of the location (bus $h$) and direction (towards $j$) observed.  It is, in turn, only dependent on the CFs of the current injected at their terminals.  

In the remainder of this paper, we work with the model described in this section and therefore the coherency function considered is (\ref{eq:coherency_function}), hereinafter denoted simply as $\bar{\epsilon}_{1,2}$.  Consequently, devices $d_{\rm{1}}$ and $d_{\rm{2}}$ are perfectly coherent if:
\begin{equation}\label{eq:coherency_condition}
    \boxed{\bar{\epsilon}_{1,2}=0\Leftrightarrow\etaidev{d_{\rm 1}}=\etaidev{d_{\rm 2}}}
\end{equation}

Note that the coherency function is always well-defined for shunt-connected devices, as it does not rely on  anything specific of the device model.  In the following section, we impose (\ref{eq:coherency_condition}) on a variety of common power system device models to study the specific conditions they must satisfy to be coherent with another instance of the same type.  Before presenting this analysis, we introduce some useful equations that are direct consequences of (\ref{eq:coherency_condition}).

First, (\ref{eq:coherency_condition}) implies that the currents satisfy:
\begin{equation}\label{eq:coherency_condition2}
    \bar{\imath}_1=\bar{\imath}_2\,\bar{k}\,,
\end{equation}
where $\bar{k}=k\,e^{\jmath\,\gamma}$ is constant.  This condition can be referred to as the currents being complex-proportional.  From (\ref{eq:coherency_condition2}) it follows immediately that:
\begin{equation}\label{eq:i2_prop}
    \imath^{2}_{1}=\imath^{2}_{2}\,k^2\,.
\end{equation}
In addition, if the devices are connected to the same bus, it can be demonstrated that:
\begin{equation}\label{eq:s_complexprop}
    \bar{s}_1=\bar{s}_2\,\bar{k}^{*}\,,
\end{equation}
which, in other words, implies that the complex powers are also complex-proportional by a factor $\bar{k}^{*}$.

\section{Coherency among specific devices}
\label{sec:coherency_devices}

In this section, we present a systematic procedure to study the conditions that two instances of specific devices must satisfy to be coherent.  The methodology consists of starting from the set of DAEs comprising a device model, finding the coherency function using (\ref{eq:coherency_function}), and imposing the condition for perfect coherency according to (\ref{eq:coherency_condition}).  The goal is to find the conditions on their variables and/or parameters required for coherency.  We examine a variety of relevant devices, including synchronous machines, converters, and loads.

\subsection{Synchronous machines}
\label{sec_SM}
Consider the lossless classical model of synchronous machines:
\begin{align}
    \dot{\delta}&=\Omega_{\rm b}(\omega_{\rm r}-1)\,,\\
    M\dot{\omega}_{\rm r}&=p_{\rm m}-p_{\rm e}-D(\omega_{\rm r}-1)\,,\label{eq:syn2_swing}
\end{align}
along with the algebraic equations:
\begin{align}
    0&=\jmath\, x_{\rm d}'\,\bar{\imath}-\bar{E}+\bar{v}\,,\\
    0&=\bar{E}\exp(-\jmath\,(\delta-\pi/2))-\jmath\, e'_{\rm q}\,,\\
    0&=\Re\{\bar{E}\,\bar{\imath}^{*}\}-p_{\rm e}\,,
\end{align}
where $\bar{\imath}$ and $\bar{v}$ are the net current injected and the voltage at terminals, respectively.  The rest of the symbols have the usual meanings.  We have previously studied this model in \cite{cfd}, where we found an expression for the CF of the current injected as follows:
\begin{equation}\label{eq:syn2_etaidev}
    \etaidev{}=\frac{\bar{s}}{\jmath\,x_{\rm d}'\,\imath^{2}}\left(\jmath\,\omega_{\rm r}-\overline{\eta^{v}}\right)+\jmath\,\omega_{\rm r}\,.
\end{equation}

Consider two instances of SMs denoted using subscripts 1 and 2.  The coherency function between them is found according to (\ref{eq:coherency_function}):
\begin{equation}\label{eq:syn2_cohfunc}\boxed{
\begin{aligned}
    \bar{\epsilon}_{1,2}&=\jmath\, (\omega_{\rm r1}-\omega_{\rm r2})+\\
    &+\frac{\bar{s}_1}{\jmath\,x'_{\rm d1}\imath_1^{2}}\left(\jmath\,\omega_{\rm r1}-\overline{\eta^{v}_1}\right)-\frac{\bar{s}_2}{\jmath\,x'_{\rm d2}\imath_2^{2}}\left(\jmath\,\omega_{\rm r2}-\overline{\eta^{v}_2}\right)\,.
\end{aligned}}
\end{equation}

By imposing (\ref{eq:coherency_condition}), two perfectly coherent SMs satisfy the following equation:
\begin{equation}
    \omega_{\rm r1}+\frac{\bar{s}_1}{x_{\rm d1}'\,\imath_{1}^{2}}\left(\overline{\eta^{v}_1}-\jmath\,\omega_{\rm r1}\right) = \omega_{\rm r2}+\frac{\bar{s}_2}{x_{\rm d2}'\,\imath_{2}^{2}}\left(\overline{\eta^{v}_2}-\jmath\,\omega_{\rm r2}\right)\,,\label{eq:syn2_0}
\end{equation}
which poses a condition on the internal behavior of the SMs, being $\omega_{\rm r}$ and $x'_{\rm d}$ the key internal variables and key parameters for coherency, respectively.  The effect of the `external' system, i.e., the rest of the network, is captured on the CF of the terminal voltage, $\overline{\eta^{v}}$.

In transient conditions, although (\ref{eq:syn2_0}) might be satisfied in different ways, below we present an intuitive set of conditions that are sufficient for perfect coherency, which allow us to deduce some insights about how SMs become coherent.  For instance, the following two conditions imply (\ref{eq:syn2_0}):
\begin{equation}\label{eq:syn2_01}
    \omega_{\rm r1}=\omega_{\rm r2}\, \land \,\frac{\bar{s}_1}{x_{\rm d1}'\,\imath_{1}^{2}}\left(\overline{\eta^{v}_1}-\jmath\right) = \frac{\bar{s}_2}{x_{\rm d2}'\,\imath_{2}^{2}}\left(\overline{\eta^{v}_2}-\jmath\right)\,.
\end{equation}

For simplicity, consider two machines connected at the same terminal bus.  In this situation, $\overline{\eta^{v}_1}=\overline{\eta^{v}_2}$, and therefore the second condition in (\ref{eq:syn2_01}) becomes:
\begin{equation}\label{eq:syn2_1}
   \frac{\bar{s}_1}{x_{\rm d1}'\,\imath_{1}^{2}}= \frac{\bar{s}_2}{x_{\rm d2}'\,\imath_{2}^{2}}\,.
\end{equation}

Replacing (\ref{eq:i2_prop}) and (\ref{eq:s_complexprop}) in (\ref{eq:syn2_1}) yields:
 \begin{align}
    \frac{\bar{s}_2\,\bar{k}^{*}}{x_{\rm d1}'\,\imath^{2}_{2}\,k^2}= \frac{\bar{s}_2}{x_{\rm d2}'\,\imath_{2}^{2}}\quad
    \Leftrightarrow\quad\frac{x_{\rm d1}'}{x_{\rm d2}'}=\frac{1}{\bar{k}}\,.
\end{align}

Hence:
\begin{equation}\label{eq:syn2_t0}
    \frac{x_{\rm d1}'}{x_{\rm d2}'}=\frac{\bar{\imath}_{2}}{\bar{\imath}_1}=\frac{\bar{\imath}_2(t_0)}{\bar{\imath}_1(t_0)}\,.
\end{equation}

Therefore, the ratio between the transient reactances of the machines shall be equal to the inverse ratio between their complex current injected.  Since the right-hand side is complex, (\ref{eq:syn2_t0}) is only satisfied if the machines have equal power factor.  This also implies that $\bar{k}$ is actually real, i.e.  $\bar{k}=k$.

Equation (\ref{eq:syn2_t0}) along with the first condition in (\ref{eq:syn2_0}) ($\omega_{\rm r1}=\omega_{\rm r2}$) form a sufficient set of conditions for two SMs connected at the same bus for being perfectly coherent.  While the former poses a condition on the parameters of the machines, the latter does so over state variables.  However, the latter has also implications over the parameters of the machine, which can be derived by taking its time derivative and recalling (\ref{eq:syn2_swing}):
\begin{align}
    \dot{\omega}_{\rm r1} =\dot{\omega}_{\rm r2} \quad 
    \Rightarrow \quad \frac{p_{\rm m1}-p_{\rm e1}}{M_1}=\frac{p_{\rm m2}-p_{\rm e2}}{M_2}\,,\label{eq:syn2_3}
\end{align}   
%
where damping has been neglected for simplicity.  Since the machines are lossless, $p_{\rm e}=p=\Re\{\bar{v}\,\bar{\imath}^{*}\}$.  Therefore $p_{\rm e1}$ can be written in terms of $p_{\rm e2}$ as follows:
\begin{align}
    p_{\rm e1}=\Re\{\bar{v}\,\bar{\imath}_1^{*}\}=\,\Re\{\bar{v}\,\bar{\imath}_2^{*}\,k\}\quad
    \Rightarrow \quad p_{\rm e1}=k\,p_{\rm e2}\,.  \label{eq:syn2_4}
\end{align}

Using (\ref{eq:syn2_4}) in (\ref{eq:syn2_3}):
\begin{equation}
    \frac{p_{\rm m1}-k\,p_{\rm e2}}{M_1}=\frac{p_{\rm m2}-p_{\rm e2}}{M_2}\,.\label{eq:syn2_5}
\end{equation}

Evaluating (\ref{eq:syn2_5}) at $t=t_0^{+}$, $p_{\rm m1}=k\,p_{\rm m2}$:
\begin{align}
    \frac{k\,p_{\rm m2}-k\,p_{\rm e2}}{M_1}=\frac{p_{\rm m2}-p_{\rm e2}}{M_2}\,\quad
    \Rightarrow \quad\frac{M_1}{M_2}=k\,.\label{eq:syn2_t02}
\end{align}

Equations (\ref{eq:syn2_t0}) and (\ref{eq:syn2_t02}) are the sought set of conditions on the parameters of two parallel lossless classical SMs to be perfectly coherent, and can be written together as follows:
\begin{equation}
    \boxed{
    \frac{x'_{\rm d1}}{x'_{\rm d2}}=\frac{M_2}{M_1}=\frac{\imath_{2}(t_0)}{\imath_{1}(t_0)}
    }\label{eq_coh_cond_SM}
\end{equation}

This analytical result is consistent with classical knowledge on coherency analysis of SMs.  In fact, an equivalent condition was for example early noticed in \cite{machowski}.  However, the difference is that we start from a general definition, over which the same set of conditions is recovered after taking a series of simplifications, yet the definition of coherency itself does not rely on these simplifications.  In turn, (\ref{eq:syn2_cohfunc}) is the general coherency function for a pair of SMs considering their classical model.
\subsection{ZIP Loads}
Consider the standard ZIP model of loads as follows:
\begin{align}
    p &= p_0\,(k_{\rm pp}+k_{\rm \imath p}v+k_{\rm zp}v^{2})\,,\\
    q &= q_0\,(k_{\rm pq}+k_{\rm \imath q}v+k_{\rm zq}v^{2})\,,
\end{align}
where $k_{ij}\in \left[0, 1\right]$, $i\in\{\rm p, \imath, z\}$, $j\in\{\rm p, q\}$, such that: 
\begin{align}
    k_{\rm pp}+k_{\rm \imath p}+k_{\rm zp}=1\,\quad \land\quad
    k_{\rm pq}+k_{\rm \imath q}+k_{\rm zq}=1\,.
\end{align}

\subsubsection{Constant impedance loads}
For a constant impedance load (Z-load), namely $k_{\rm zp}=k_{\rm zq}=1$, the CF of the current injected by the load is \cite{cfd}:
\begin{equation}
    \etaidev{}=\overline{\eta^{v}}\,,
\end{equation}
where $\overline{\eta^{v}}$ is the CF of the terminal voltage.

Consider two instances of Z-loads denoted using subscripts 1 and 2.  The coherency function between them is found according to (\ref{eq:coherency_function}):
\begin{equation}\boxed{
    \bar{\epsilon}_{1,2}=\overline{\eta^{v}_1}-\overline{\eta^{v}_2}}
\end{equation}

By imposing (\ref{eq:coherency_condition}) we find that two perfectly coherent Z-loads satisfy the following condition:
\begin{equation}
    \overline{\eta^{v}_1}=\overline{\eta^{v}_2}\,.
\end{equation}

Therefore, the loads are perfectly coherent as long as the CF of their terminal voltage is equal.  In transient conditions, this is generally not satisfied exactly unless the two loads are connected at the same terminal bus.  However, nearby buses will tend to be coherent as the voltage tends to be close.  The coherency analysis shows that coherency among Z-loads is entirely dependent on the network characteristics and the behavior of `the rest of the system', as there is no constraint on the load's internal variables or parameters.  This result is in accordance with their entirely passive nature.

\subsubsection{Constant power loads}
For a constant power load (S-load), namely $k_{\rm pp}=k_{\rm pq}=1$, the CF of the current injected by the load is \cite{cfd}:
\begin{equation}
    \etaidev{}=-\overline{\eta^{v}}^{*},
\end{equation}
where $^*$ denotes the conjugate operator.

Despite the difference in the sign and the conjugate operation, the same discussion given before for the Z-load applies to the S-load.  In general, coherency among any type of ZIP loads will depend on the parameters of the grid and the behavior of the rest of devices.  Loads electrically close will tend to behave coherently given the CF of their terminal voltages will exhibit a very similar trajectory in transient conditions.
\subsection{Inverter-based resources (IBRs)}
Consider a standard IBR model of a controlled voltage source behind a filter.  The modulated signal is $\bar{m}$, and $\bar{z}_{\rm f}$, $\bar{y}_{\rm f}$ are the output filter series impedance and shunt conductance, respectively.  The set of algebraic equations describing such model is as follows:
\begin{align}
    \bar{v}&=\bar{e}-\bar{z}_{\rm f}(\bar{\imath}+\bar{y}_{\rm f}\bar{v})\,,\\
    \bar{e}&=\bar{m} \, v_{\rm dc0}\,,\\
    \bar{s}&=\bar{v}\,\bar{\imath}^{*}\,.
\end{align}

The CF of the current injected by this device is:
\begin{equation}\label{eq:gfl_etaidev}
    \etaidev=\frac{\bar{s}}{\bar{z}_{\rm f}\,\imath^2}(1+\bar{z}_{\rm f}\,\bar{y}_{\rm f})\left(\overline{\eta^{e}}-\overline{\eta^{v}}\right)+\overline{\eta^{e}}\,.
\end{equation}

Equation (\ref{eq:gfl_etaidev}) has a very similar structure to the expression for the classical model of a SM (equation (\ref{eq:syn2_etaidev})).  In fact, the latter is recovered from the former after replacing $\bar{z}_{\rm f}=\jmath\,x'_{\rm d}$, $\bar{y}_{\rm f}=0$, and $\overline{\eta^{e}}=\jmath\,\omega_{\rm r}$.  Equation (\ref{eq:gfl_etaidev}) is actually a general result for any device with an equivalent circuit representation of a voltage source $\bar{e}$ behind a series impedance $\bar{z}_{\rm f}$ and a shunt admittance $\bar{y}_{\rm f}$ connected at the terminal bus.  What remains to complete the model are specific expressions for $\overline{\eta^{e}}$, which depend on the internal control loops composing the dynamic model of the converter.  We provide examples for typical converter models below.
\subsubsection{Grid-following (GFL) converters}
Consider a standard GFL converter model with a synchronous reference Phase Locked Loop (SRF-PLL), and PI controllers for the internal current control loops with fixed references.  The set of differential equations is as follows:
\begin{align}
    \dot{\overline{x}}&=K_{\rm i}(\overline{\imath}_{\rm ref}-\overline{\imath}_{\rm m})\,,\\
    T_{\rm m }\dot{\overline{\imath}}_{\rm m}&=\overline{\imath}-\overline{\imath}_{\rm m}\,,\\
    \dot{x}_{\rm pll}&=K_{\rm i_{\rm pll}}v_{\rm q}\,,\\
    \dot{\tilde{\theta}}&=\Delta\omega_{\rm pll}\,,
\end{align}
%
along with the algebraic equations:
\begin{align}
    \bar{m}&=\bar{x}+K_{\rm p}(\bar{\imath}_{\rm ref}-\bar{\imath}_{\rm m})\,,\\
    \Delta\omega_{\rm pll}&=K_{\rm p_{\rm pll}}v_{\rm q}+x_{\rm pll}\, ,\\
    \tilde{\omega}&=\Delta\omega_{\rm pll}+\omega_{\rm ref}\, ,\\
  \bar{v} &= v_{\rm d} + \jmath \, v_{\rm q} = v\cos(\theta-\tilde{\theta}) +
            \jmath\, v\sin(\theta-\tilde{\theta}) \, ,\\
  \bar{\imath} &= \imath_{\rm d} + \jmath \, \imath_{\rm q} =
                 \imath\cos(\theta-\tilde{\theta}) + \jmath\,\imath\sin(\theta-\tilde{\theta})
                 \,.
\end{align}

For this model, the CF of the internal voltage source is:
\begin{equation}
    \overline{\eta^{e}}=\frac{\dot{m}}{m}+\jmath\,(\dot{\alpha}+\tilde{\omega})\,,
\end{equation}
where $m$ is the magnitude of $\bar{m}$, and $\alpha$ its angle in the converter internal $\rm dq$ coordinates.

Consider two instances of GFLs denoted using subscripts 1 and 2.  The coherency function between them is found according to (\ref{eq:coherency_function}):
\begin{equation}\label{eq:gfl_cohfunc}\boxed{
\begin{aligned}
    \bar{\epsilon}_{1, 2}&=\frac{\dot{m}_1}{m_1}-\frac{\dot{m}_2}{m_2}+\jmath\,\left(\dot{\alpha}_1+\tilde{\omega}_1-\dot{\alpha}_2-\tilde{\omega}_2\right)+\\
    &+\frac{\bar{s}_1}{\bar{z}_{\rm f1}\,\imath^2_1}\left(\frac{\dot{m}_1}{m_1}+\jmath\,(\dot{\alpha}_1+\tilde{\omega}_1)-\overline{\eta^{v}_1}\right)+\\
    &-\frac{\bar{s}_2}{\bar{z}_{\rm f2}\,\imath^2_2}\left(\frac{\dot{m}_2}{m_2}+\jmath\,(\dot{\alpha}_2+\tilde{\omega}_2)-\overline{\eta^{v}_2}\right)
\end{aligned}}
\end{equation}

As evidenced by (\ref{eq:gfl_cohfunc}), key variables for coherency between GFLs are $\tilde{\omega}$, namely, the frequency estimation given by the PLL for synchronization, and the derivatives of $m$ and $\alpha$, which, ultimately, depend on the internal current control loops of the converter.  Once again, the effect of the network and the rest of the system is captured through $\overline{\eta^{v}}$.

\subsubsection{Grid-forming (GFM) converters}
Consider a standard GFM converter model with a power-frequency droop for synchronization and a PI control loop for the voltage magnitude.  The set of differential equations is as follows:
\begin{align}
    \dot{e}&=K_{\rm i}(v_{\rm ref}-v_{\rm m})-\frac{K_{\rm p}}{T_{\rm v}}(v_{\rm m}-v) \, , \\
    \dot{\delta}&=\Omega_b(\omega_{\rm gfm}-1)\, ,
\end{align}
along with the algebraic equations:
\begin{align}
    \omega_{\rm gfm}&=m_{\rm p}(p_{\rm ref}-p_{\rm m})+1\, ,\\
    \bar{e}&=\bar{m} \, v_{\rm dc0}\,.
\end{align}

For this model, the CF of the internal voltage source is:
\begin{equation}
    \overline{\eta^{e}}=\frac{\dot{e}}{e}+\jmath\,\omega_{\rm gfm}\,.
\end{equation}

Consider two instances of GFMs denoted using subscripts 1 and 2.  The coherency function between them is found according to (\ref{eq:coherency_function}):
\begin{equation}\label{eq:gfm_cohfunc}\boxed{
\begin{aligned}
    \bar{\epsilon}_{1, 2}&=\frac{\dot{e}_1}{e_1}-\frac{\dot{e}_2}{e_2}+\jmath\,\left(\omega_{\rm gfm1}-\omega_{\rm gfm2}\right)+\\
    &+\frac{\bar{s}_1}{\bar{z}_{\rm f1}\,\imath^2_1}\left(\frac{\dot{e}_1}{e_1}+\jmath\,\omega_{\rm gfm1}-\overline{\eta^{v}_1}\right)+\\
    &-\frac{\bar{s}_2}{\bar{z}_{\rm f2}\,\imath^2_2}\left(\frac{\dot{e}_2}{e_2}+\jmath\,\omega_{\rm gfm2}-\overline{\eta^{v}_2}\right)
\end{aligned}}
\end{equation}

The GFMs coherency function has an analogous structure to that of GFLs, being now the key variables the magnitude of the internal voltage $e$ and the internal frequency set by the converter for the power-synchronization mechanism $\omega_{\rm gfm}$.

\section{Case Study}
\label{sec:casestudy}

In this section, we evaluate coherency among devices using two case studies.  The first case consists of a single node where two SMs and a constant impedance load are connected.  This setup is used to study the fundamental concepts and relationships of coherency between two well-known second-order model of SMs.  The second case involves the IEEE 39-bus system, which has been historically used for studying coherency among SMs.  Several methods for identifying coherent generators have been tested in this system, making the groups of machines that swing together well-known and accepted, an ideal feature for a case study to use as a reference base example.  A modified version of this system is also included, in which some SMs are replaced by GFL and GFM converters to evaluate coherency across different technologies.  All simulations are performed using the software Dome \cite{dome}.

\subsection{Two-machine system}
\begin{figure}[hbtp]
\vspace{-.25cm}
    \centering
    \includegraphics[width=0.9\linewidth]{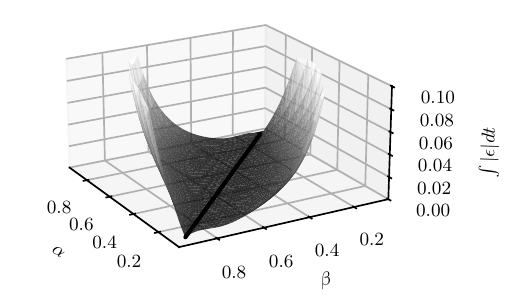}
    \caption{Two-machine system - Time integral of the magnitude of the coherency function as a function of $\alpha$ and $\beta$.}
    \label{fig:epsilon_vs_alpha_beta}
\end{figure}

A constant impedance load and two second-order synchronous machines (SMs), described in \ref{sec_SM}, are connected in parallel at a single node.  Coherency between the SMs is studied through time-domain simulations for different inertia constants, transient reactances, and currents, and evaluated using the coherency condition in (\ref{eq_coh_cond_SM}).  Sensitivity is studied through the parameters $\alpha = M_1 / (M_1 + M_2)$ and $\beta = x_{\rm d1}' / (x_{\rm d1}' + x_{\rm d2}')$, assuming initial dispatch proportional to inertia ($M_1 / M_2 = \imath_1 / \imath_2$), a total inertia $M_1+M_2=10$ s, total current $\imath_1+\imath_2=1$ pu, and a total series reactance $x_{\rm d1}' + x_{\rm d2}'=0.1$ pu.  Figure~\ref{fig:epsilon_vs_alpha_beta} shows the time integral of the magnitude of the coherency function (\ref{eq:gendef}) as a function of $\alpha$ and $\beta$.  Each simulation includes a small imbalance perturbation equivalent to 10\% load increase at $t=1$ s, removed after 10 ms to restore the initial condition.

Perfect coherency ($\int|\epsilon|,dt=0$) is observed when $\alpha = 1 - \beta$ (highlighted with a bold line), consistent with the condition in (\ref{eq_coh_cond_SM}).  In this case, both SMs behave coherently, and their dynamics can be represented by the states of a single machine, with their powers being proportional throughout the simulation.  For all other parameter values, power oscillations arise between the SMs, and the coherency function increases as $\alpha$ and $\beta$ deviate from the line $\alpha = 1 - \beta$.

\subsection{IEEE 39-Bus system}
\begin{figure}[hbtp]
  \centering
  {%
    \includegraphics[width=0.75\linewidth]{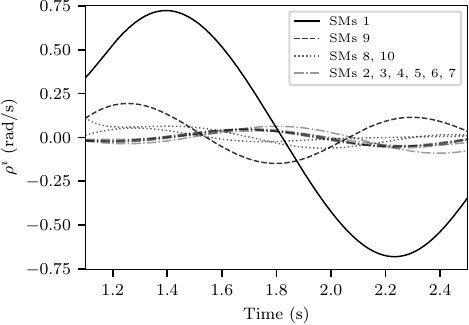}%
    }
  \\
  {%
    \includegraphics[width=0.75\linewidth]{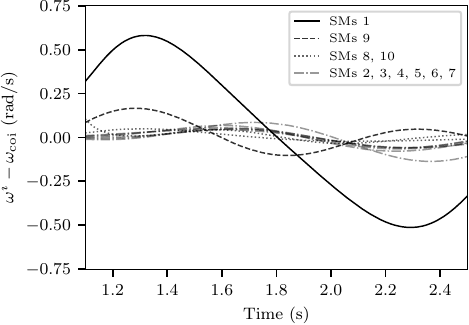}%
   }
  \caption{Real (upper panel) and imaginary (lower panel) parts of the CF of the SM currents.}
  \label{fig:etai_ieee39}
\end{figure}

The system is parametrized and initialized according to \cite{IEEE39}.  A time-domain simulation is conducted for a partial disconnection of 100 MW of load at bus 28 at $t=1$ s.  The trajectories of the CF of the current of the SMs are computed as in \cite{cfd} and shown in Fig.  \ref{fig:etai_ieee39} for a time window following the event, where the real and imaginary parts of the CF are denoted as $\rho^{\imath}$ and $\omega^{\imath}$, respectively.

\begin{figure}[hbtp]
    \centering
    \includegraphics[width=0.77\linewidth]{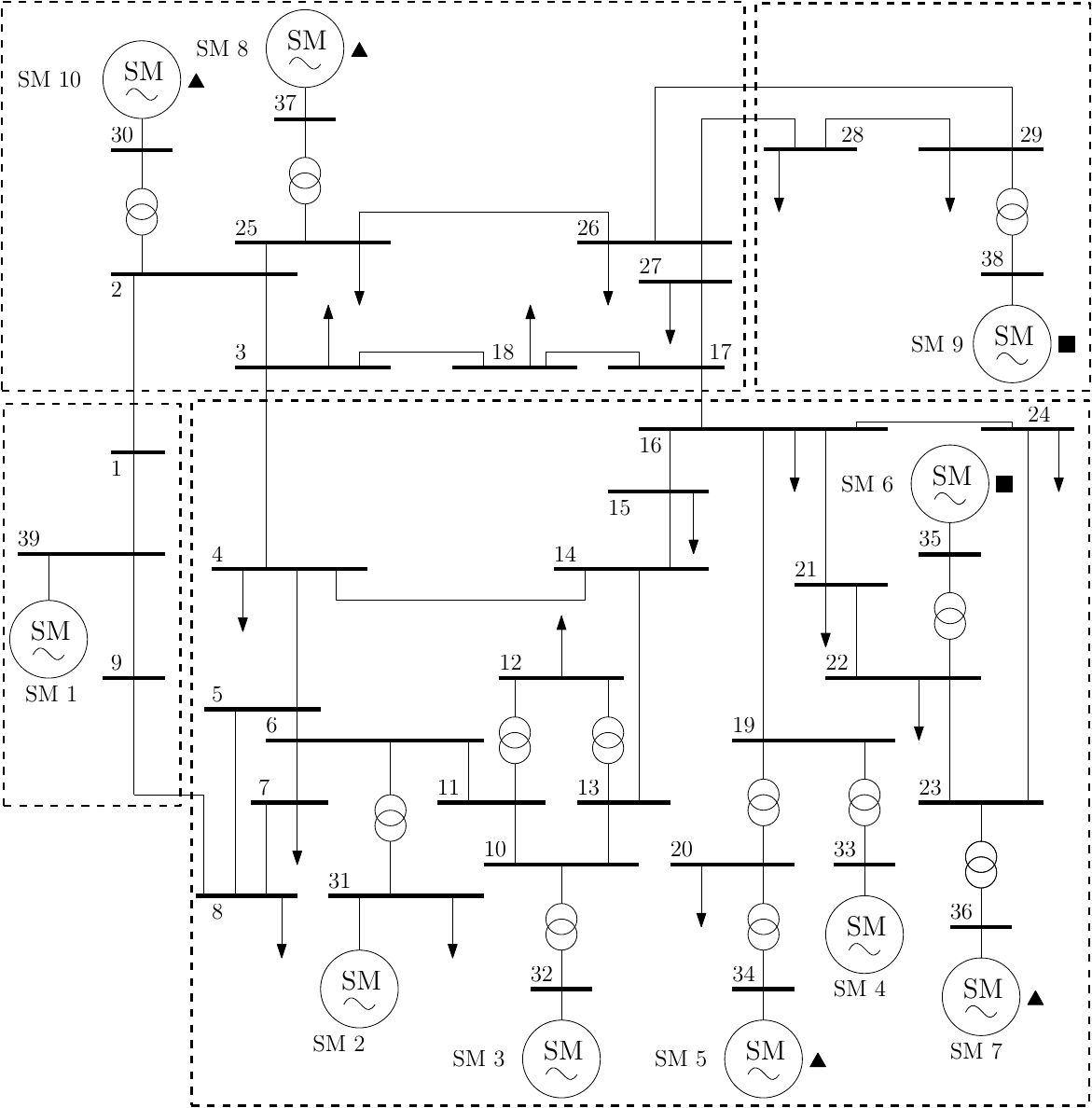}
    \caption{Resulting coherent groups of SMs for a four-area division.  Triangle $\left(\blacktriangle\right)$ and rectangles $\left(\blacksquare\right)$ markings identify the generators replaced by GFL-IBRs and GFM-IBRs in the next case study.}
    \label{fig:ieee39results}
\end{figure}

A standard average linkage clustering method is implemented using the time integral of the magnitude of the coherency function (\ref{eq:gendef}) as the distance metric.  For instance, in the case of a four-area division, the results are illustrated in Fig.  \ref{fig:ieee39results}.  Generators 2–7 form a coherent group, as do SMs 8 and 10, while SM 1 and 9 remain isolated.  These results are consistent with classical knowledge on this system, very similar, for example, to the results on \cite{chow, yusof}.

\subsection{Modified IEEE 39-Bus system}

Coherency is evaluated in a modified version of the IEEE 39-Bus system, now composed of an heterogeneous mix of devices.  To do so, some SMs are replaced by IBRs using the models described in Section \ref{sec:coherency_devices}.  The generators in Fig.  \ref{fig:ieee39results} marked with triangles $\left(\blacktriangle\right)$ are replaced by GFL-IBRs, and those marked with rectangles $\left(\blacksquare\right)$ by GFM-IBRs, whereas loads are still considered constant impedances.  

\begin{figure}[hbtp]
  \centering
  {%
    \includegraphics[width=0.75\linewidth]{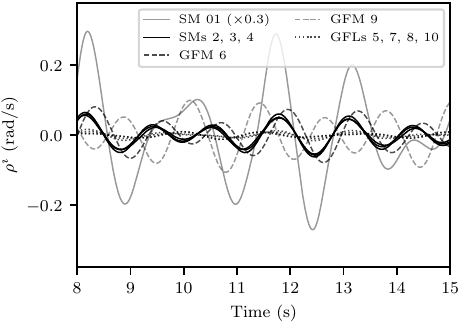}%
    }
  \\
  {%
    \includegraphics[width=0.75\linewidth]{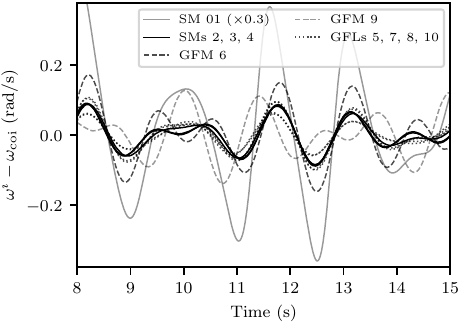}%
    }
  \caption{Real (upper panel) and imaginary (lower panel) parts of the CF of the device currents.}
  \label{fig:etai_MOD_ieee39}
\end{figure}

A time-domain simulation is carried out for the same disturbance as in the previous case.  The trajectories of the CF of the current injected by the devices are shown in Fig.~\ref{fig:etai_MOD_ieee39}.
Focusing on the real part of the CF ($\rho^{\imath}$), it is observed that SMs 2, 3 and 4 oscillate together, while SM 1 remains alone as in the previous case.  
Interestingly, the four GFL-IBRs are coherent, even though they are not necessarily close to each other in the network (see for example, GFL 5 at bus 34 and GFL 10 at bus 30).  These devices tend to follow an averaged response of the system.  Regarding GFM-IBRs, the one connected at bus 38 (GFM 9) oscillates alone, as do GFM 6.  However, the latter tends to be closer to the group of SMs 2, 3 and 4.  By observing the imaginary part of the CF ($\omega^{\imath}$), there is a broader manifestation of coherency across technologies, since the group of SMs 2, 3 and 4 is coherent with the four GFLs and GFM 6, while SM1 and GFM 9 remain oscillating individually.

As our definition is independent from devices' technology, coherency analysis can be conducted in the same way as in the previous example.  
In particular, for this case study, the same clustering implementation considering a four-area division in the modified system gives the following results: SM 1 and GFM 9 remain isolated, the four GFLs form a coherent group, as do SMs 2, 3, 4 alongside GFM 6.

The manifestation of coherency among devices proves to be a combination of two factors: their technology (same-kind models react similarly to perturbations), and their location (close devices tend to be coherent).  In this example, the former factor seems to mostly impact $\rho^{\imath}$, and the latter $\omega^{\imath}$.  Of course, further investigation is needed to fully reveal the role of $\rho^{\imath}$ and $\omega^{\imath}$ in terms of the particular information they carry.  

\section{Conclusion}
\label{sec:conclusion}

The paper proposes a novel definition of coherency among power system devices as the difference between the CF of the net current injected by devices.  This definition has various relevant implications.
First, the independence of coherency from the observer.  In other words, if two devices are perfectly coherent during a transient, they are so from the point of view of any bus and variable of interest.
Then, our definition is model agnostic, making it suitable for evaluating coherency among shunt-devices of any kind.  It is therefore a generalization of coherency of power system devices.  In fact, if applied to synchronous machines under usual assumptions, same results and conclusions are recovered from those obtained from conventional coherency analysis.
Finally, our definition is based on electrical variables easily measurable at device terminals.  

In future work, we will exploit the latter feature to develop online security assessment tools, and for control applications, e.g., configuring coherency as a control objective.  Besides, we believe our definition is an important step towards a more general and systematic framework for the aggregation of an heterogeneous mix of devices, which benefits model reduction techniques, wide-area control zones identification, among other applications for modern power systems.



\end{document}